\documentclass[12pt,aps,prb,preprint]{revtex4}

\usepackage{graphics} \usepackage{graphicx}\usepackage{amsmath}

\newcommand{\str}[1]{$\{ #1 \}$} \newcommand{\av}[1]{\langle #1
\rangle} 

\begin{document}

\title{ The Influence of Liquid Structure on the Thermodynamics of Freezing}

\author{Pierre Ronceray} \affiliation{\textit{Laboratoire de Physique Th\'eorique et Mod\`eles Statistiques, Universit\'e Paris-Sud, B\^at. 100, 91405 Orsay
Cedex, France}\\
  \textit{D\'epartement de Physique, \'Ecole Normale Sup\'erieure,
    75005 Paris}}

\author{Peter Harrowell} \affiliation{\textit{School of Chemistry,
University of Sydney, Sydney N.S.W. 2006, Australia}}

\date{\today} 

\begin{abstract}
  The ordering transitions of a 2D lattice liquid characterized by a
  single favoured local structure (FLS) are studied using Monte Carlo
  simulations. All eight distinct geometries for the FLS are
  considered and we find a variety of ordering transitions - first
  order, continuous and multi-step transitions. Using an
  entropy-energy representation of the freezing transition we resolve
  the dual influence of the local structure on the ordering
  transition, \emph{i.e}. via the energy of the crystal and the
  entropy cost of structure in the liquid. The generality of this approach in demonstrated in an analysis of the influence on the tetrahedrality of a modified silicon model and its freezing point.

\end{abstract}

\maketitle

\section{Introduction}
\label{intro}

As a liquid is cooled down, its molecules will seek energetically
favorable arrangements at the microscopic scale. It is reasonable to
presume that among these local structural arrangements that are found
to be particularly stable in the liquids will be the ones repeated
periodically in the crystals. It is difficult to see how a liquid
could freeze if this was not so. Let us imagine that the crystal is
assembled so as to maximise the density of just a single local
structure. It would follow that a minimal condition on any Hamiltonian
used to model both the liquid and crystal phases is that it must
stabilize this particular favoured local structure. The simplest such
model would be the one that {\it{only}} stabilized this one local
structure. This is a simpler picture than is usually invoked for
liquid structure. There is, for example, an extended
tradition~\cite{frank,tarjus} of focussing on non-crystalline favoured
local structures in order to explain the kinetic stability of the
supercooled liquid with respect to freezing. While eminently
reasonable, the idea that a competing structure is {\it necessary} to
explain liquid metastability does imply that a single stable local
structure could not provide this stability, irrespective of its
geometry. Providing a test of this implication is a task for which our
minimal liquid model is ideally suited. In this paper we shall explore
the consequences of the minimal model on the liquid and solid
states. Our primary interest is to understand how the geometry of the
favoured local structure (FLS) determines the nature of the freezing
transition. We shall show that the FLS exerts its influence in
different ways in the crystal and liquid phases: determining the
lattice energy of the former and the entropy cost per FLS in the
liquid.

To address the relationship between the liquid structure and freezing,
we shall use a 2D lattice model, the Favoured Local Structure (FLS)
model, which allows us to directly select the geometry of a favoured
local structure from among the possible
geometries~\cite{ronceray_short}. Previously, we have identified the crystal ground states for each
possible choice of FLS~\cite{ronceray_short} and demonstrated that the
decrease in liquid entropy with decreasing energy is smaller for
liquids based on a low symmetry FLS as compared with those based on
high symmetry local structures~\cite{ronceray_liq}. In this paper, we present an account of the freezing transitions
described by the FLS model. In doing so we shall demonstrate how the
entropy-energy representation of the phase transitions allows us to clearly
see the dual role of the local structure in both establishing the
energy of the crystal state and the entropy cost of local ordering in
the liquid state.

The question of the relationship between the local stable structure of
a liquid and the ordering transitions of that liquid has, until
recently, been limited to considerations of specific (high symmetry)
geometries. To model the consequences of a local {\it icosahedral}
coordination, Dzugutov~\cite{dzugutov92} introduced a spherical
particle whose short range attraction was augmented with a
next-to-nearest neighbour repulsion, adjusted so as to destabilize
close packed crystals structures and enhance icosahedral coordination geometries. The Dzugutov liquid was found to crystallize into a metastable dodecagonal
quasicrystal~\cite{dzugutov93} with the equilibrium crystal phases
being body-centered cubic (bcc) at low pressures and face-centered
cubic at high pressures~\cite{roth00}.

Probably the most extensively studied
example of a liquid with a selected geometry is
silicon~\cite{silicon-gen} and its structural
analogues~\cite{molinero}. A widely used potential for silicon,
developed by Stillinger and Weber~\cite{SW}, imposes a bond angle
energy for triplets of particles to select for the tetrahedral
angle. By reducing the strength of this bond angle potential, the
local tetrahedral constraint can be continuously relaxed and
equilibrium solid phase changes from the diamond structure to
bcc~\cite{sastry}.  At the value of the bond strength corresponding to
the crossover between the two crystal phases, there is a significant
freezing point depression and an associated improvement in the glass
forming ability.

Both the Dzugutov and Stillinger-Weber potentials originated in modelling of actual atomic interactions. The recent interest in
colloidal self assembly has been driven by an increasing capacity to
experimentally manipulate the particle interactions
themselves~\cite{colloids}. This new perspective has raised some new
questions, closer in spirit to the subject of this paper. What is the
simplest set of interactions required to stabilize a given target
local structure? General aspects of this question have been addressed in the tile assembly model
of Winfree and coworkers~\cite{winfree}. A more specific example has been studied by Hormoz and Brenner~\cite{hormoz} who have looked at enhancing the frequency of a high symmetry arrangement of 8 hard spheres by adjusting the nearest neighbour interactions. Restricting themselves to
isotropic interactions, Tindemans and Mulder~\cite{mulder} explored
the possibility of designing a unique crystal groundstate by minimal
extensions of the range and selectivity of the interparticle
potential. The kinetics of assembly must also depend on the geometry of
the target structure. Wilber \emph{et al}~\cite{wilber} have
considered this problem in the context of self assembly of the
Platonic polyhedra.  Among the most extensively studied models of anisotropic colloid interactions are spheres decorated by sticky patches. A number of studies of this model have addressed the question of how
does the symmetry of the local interactions influence the phase
diagram~\cite{patch}.

The present study, while also addressing the issues concerning the
relationship between favoured local structures and the thermodynamics
of the phases that result, differs from these previous studies in that
it demonstrates how the problem of the relationship between
interactions and structure can be set aside, to permit a more
transparent and complete exploration of the consequences of a given
favoured local structure (however its stability is engineered) on the
structure and stability of the associated condensed phases -
disordered as well as ordered.

\section{The FLS Model }
\label{sec:model}

\begin{figure}[p] \centering \vspace{-1.6cm}
  \includegraphics{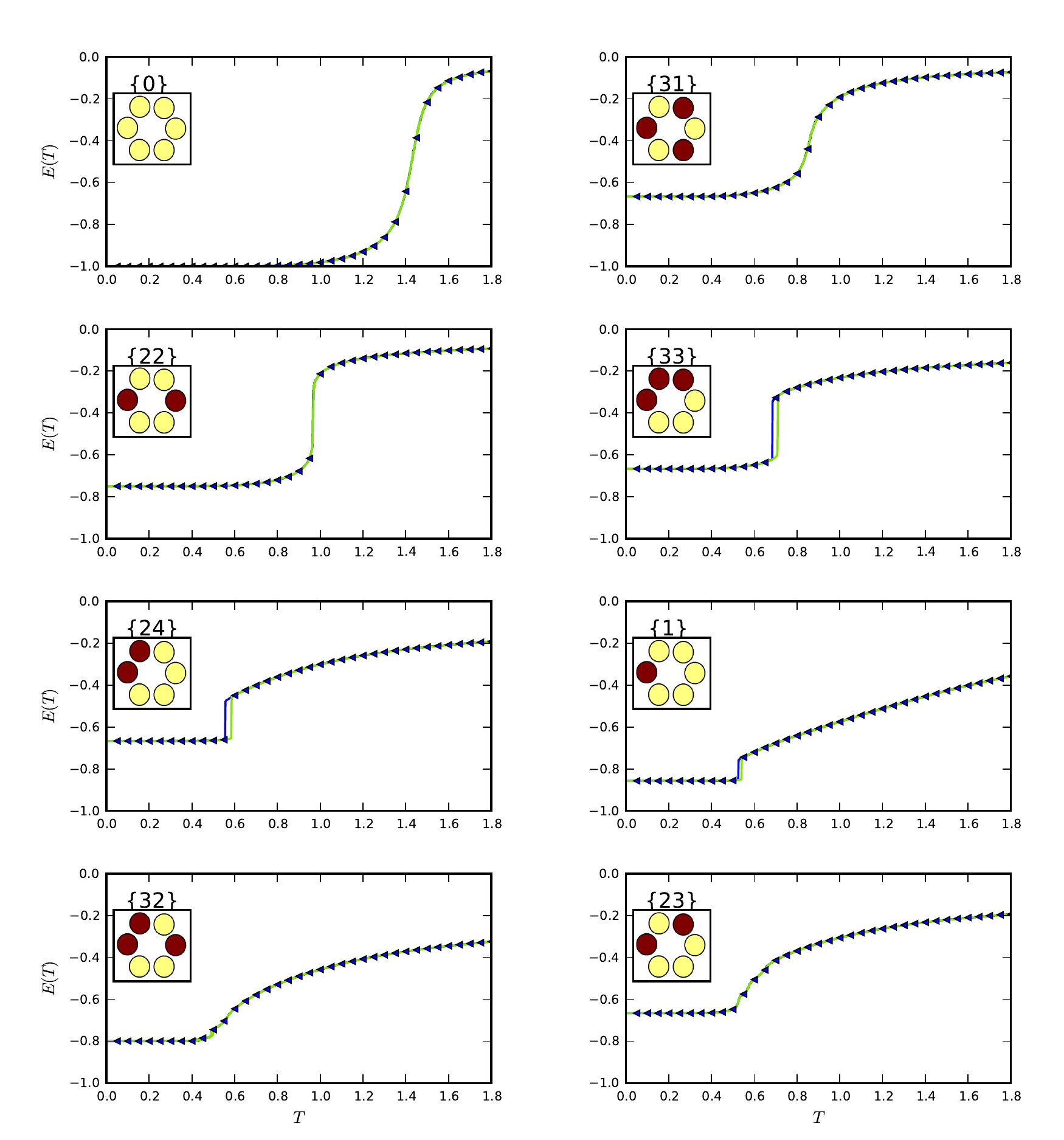}
  \caption{\label{fig:EvsT}(Color online) The average energy per site
    as a function of temperature for each FLS. The measurements were
    made in an hysteresis cycle of temperature -- cooling, in blue (dark
    gray) then heating, in green (light gray).  These curves were
    obtained with a $60\times60$ lattice and a ``cooling rate'' of
    $2.10^6$ steps of the rejection-free algorithm per site and per
    unit of temperature.}
\end{figure}

We consider a 2D triangular lattice of Ising spins (each site has a
single degree of freedom named \emph{spin value}, which can take two
values called \emph{up} or \emph{down}). These elementary variables
are taken as a minimal model for local conformational degrees of
freedom of a generic liquid. Readers should note that this model is
identical to one in which two different species A and B are
distributed on the lattice. The local environment of a given site is
defined as the spin state of its 6 nearest neighbours. In the set of
$2^6$ possible environments, there are 8 local structures that cannot
be inter-converted by rotations or spin inversion. These structures
are sketched in the insets on Figure~\ref{fig:EvsT}. The labeling
convention we adopted to design each local structure goes as follow:
the first digit is the number of down spin (dark sites on the
pictures), and the second one is the length of the longest sequence of
up spins in the structure (when there is only $0$ or $1$ down site,
this second digit is unnecessary, and associated structures will be
named simply \str{0} and \str{1} respectively). The $\{32\}$ structure
has no plane symmetry and is therefore chiral.  Here we will consider
only the case when the enantiomers are not distinguished. The chiral
case \str{32c}, where we favour only one enantiomer, involves a number
of interesting subtleties that will be addressed in a separate
paper~\cite{ronceray_chiral}.

Among these 8 distinct local structures, we select one that will be
designated as the favoured local structure (FLS). Each site whose
local environment is the FLS (to a rotation) is assigned an energy of
$-1$ ; all the others have an energy of $0$. Note that this energy is
independent of the spin on the given site, depending only on the spin
arrangement of the 6 neighbouring sites.

We evaluate the equilibrium properties for each FSL model using
Monte-Carlo sampling. We consider a system in the canonical ensemble,
at temperature $T$ (in this work we will take $k_B = 1$). For high
temperatures we have used the standard Metropolis algorithm and, at
low temperatures, the equivalent rejection-free method due to Bortz
\emph{et al.}~\cite{bortz} in which sites are organized in terms of
the energy change associated with the spin flip. We have chosen this
MC algorithm because, by using random spin flips, it retains enough of
a connection to the actual dynamics (Glauber spin dynamics, in this
case) so as to provide some indication of the existence and nature of
metastability. The price we pay for this extra information is an
uncertainty in the transition temperature equal to the hysteresis in
the transition temperature when comparing the heating and cooling
runs. As a result of the absence of any significant supercooling of
the liquid, we find the the uncertainty in the transition temperature
is sufficiently small to satisfactorily resolve the trends discussed
in this paper.

\begin{figure}[p] \centering \vspace{-1.6cm}
  \includegraphics{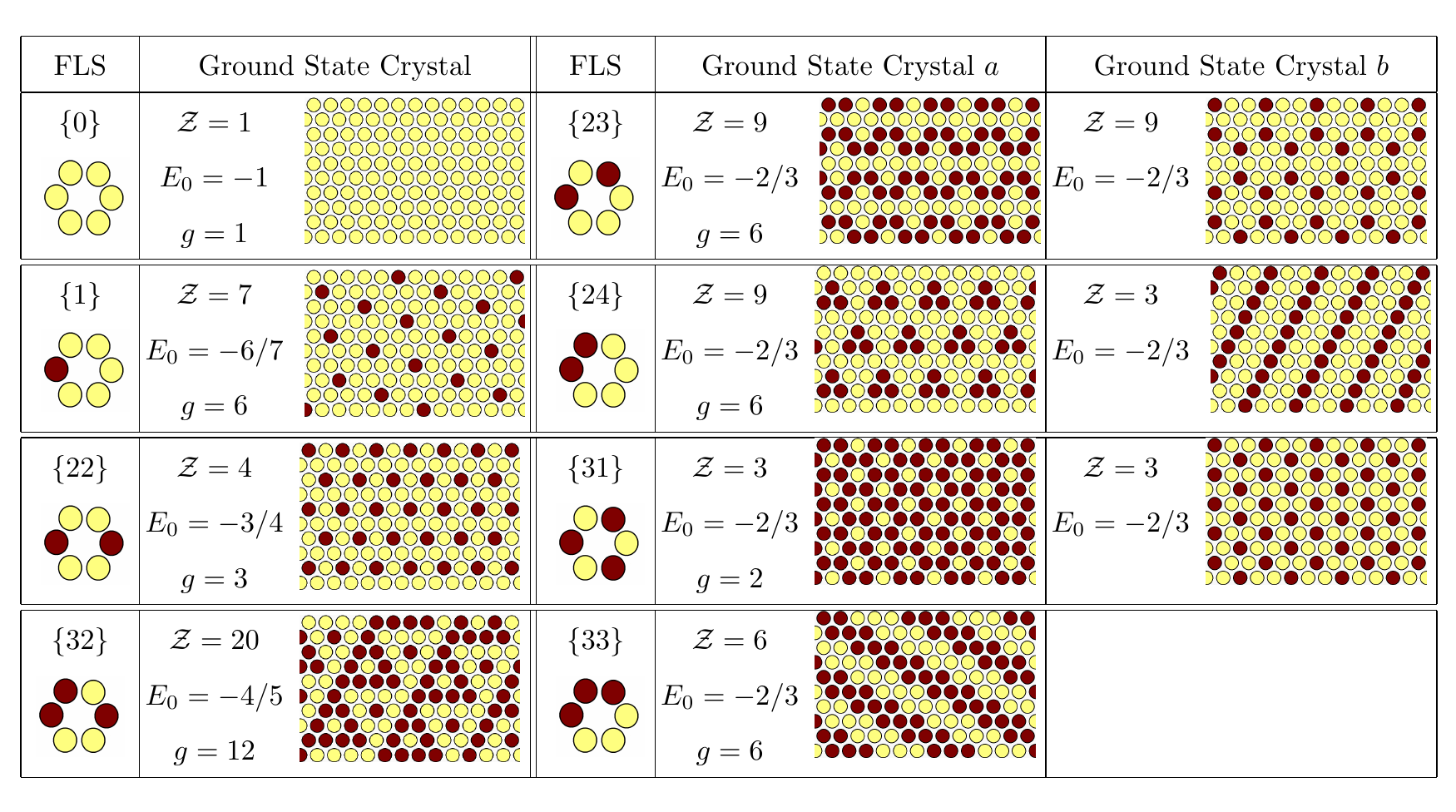}
  \caption{\label{fig:crystal}(Color online) The groundstates for the
    $8$ distinct FLS as defined in the text. The geometrical
    multiplicity $g$ of the FLS, number of sites in the unit cell
    $\mathcal{Z}$ and the energy per site $E_0$ in the groundstate are
    indicated. For the \str{23}, \str{24} and \str{31} FLS's we find
    two degenerate polymorphs. }
\end{figure}

We have previously~\cite{ronceray_short} identified the ground states
(all crystalline) for each choice of FLS.  To assist the discussion of
the various freezing transitions, we present the crystal phases for
the different FLS's in Figure~\ref{fig:crystal} along with the energy
per site $E_o$ and the number $g$ of distinct realizations of a FLS on
the lattice. Note that three FLS's have two distinct crystals with the
same energy.

\section{Crystal-Liquid Phase Transitions }
\label{sec:phase_trans}

The temperature dependence of the average energies on heating and
cooling are plotted in Figure~\ref{fig:EvsT} for each FLS. We note
that the choice of FLS can result in either first order transitions
(\str{1}, \str{24} and \str{33}), continuous transitions (\str{0},
\str{22} and \str{31}), or, in the case of \str{32} and \str{23},
multiple transitions (see Figure~\ref{fig:mult_trans} for a magnified
view in this case). In the cases where two degenerate polymorphs exist
we find that only one of the two is observed to form in the case of
\str{23} and \str{24}. The observed crystal is labeled $a$ in
Figure~\ref{fig:crystal}. In the case of \str{31}, the polymorphs are
spin-inversion symmetric and either can form spontaneously from the
liquid.

In addition to the variety of freezing transitions, the equilibrium
liquid states exhibit a considerable variation in their heat
capacities $C_{V} = \frac{dE}{dT}$ and the maximum concentration of
FLS that the liquid can accumulate before freezing. In a previous
paper~\cite{ronceray_liq} we demonstrated that the influence of the
choice of FLS on the liquid structure could be attributed to the
difference in the entropy cost per FLS. Contributions to this entropy
cost per FLS were found to come from both the symmetry of the favoured
local structure itself (the higher the symmetry, the greater the
entropy cost) and from the entropy of aggregates of FLS's.

\begin{figure}[p] \centering \vspace{-1.6cm}
  \includegraphics[width = 0.6\textwidth]{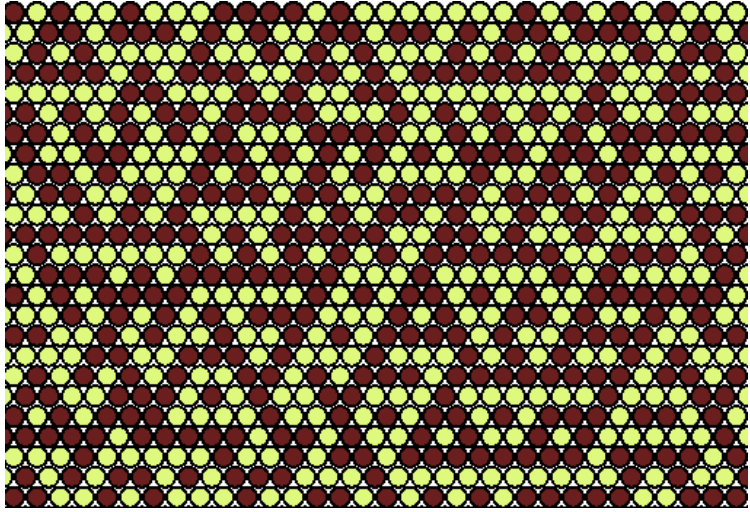}
  \caption{\label{fig:32}(Color online) Typical equilibrium spin
    configuration of the system for FLS \str{32}, at temperature
    $T=0.6$, just above the first transition. The extensive local
    structuring of the liquid appears clearly.}
\end{figure}

Not only does the decrease in symmetry of the FLS allow the liquid to
accumulate more local order before freezing, but it also tends to
result in large unit cells in the crystal~\cite{ronceray_short}. This
coincidence of effects means that large unit cells in the crystal
phase will often be associated with a liquid that can accumulate a
significant amount of local structure (i.e. 'pre-order') before
freezing and help explain why we see little difference in the rate of
crystallization of the different systems, even though the unit cell
varies across a factor of 20. To appreciate the extent of this
pre-ordering, consider the case of the \str{32} FLS which freezes into
our most complicated crystal structure (see
Figure~\ref{fig:crystal}). This freezing involves multiple transitions
(as we shall describe below) but if we select $T = 0.6$, a temperature
that lies just above the first of the ordering transitions, we find
that the local ordering has already reached roughly $75\%$ of the
final crystal order. In Figure~\ref{fig:32} we show an example
configuration in the \str{32} liquid at $T = 0.6$ in which the
entangled fragments of the crystalline order are evident.

\section{The Common Tangent Construction on the Entropy-Energy Plane }
\label{sec:common_tangent}

Two phases are in coexistence when their free energies are
equal. Under the fixed size and temperature conditions of our
simulations, the appropriate free energy is the canonical or Helmholtz
free energy $F(T)=E(T)-TS(T)$ and the freezing point $T_f$ is
identified through the resulting relation between the average energy
$E(T)$ and the entropy $S(T)$,

\begin{equation}
\label{eq:helmholtz} E_{\mathrm{crystal}}(T_{f})-T_{f}S_{\mathrm{crystal}}(T_{f})=E_{\mathrm{liquid}}(T_{f})-T_{f}S_{\mathrm{liquid}}(T_{f})
\end{equation}

\noindent For freezing transitions, a useful representation of the coexistence condition is a common
tangent construction on the entropy-energy plane. This construction
follows directly from Eq.~\ref{eq:helmholtz} and the
relation
\begin{equation}
\label{eq:partial} dS(T) =  \frac{1}{T} dE(T)
\end{equation}
This relation between canonical quantities resembles the relation by which temperature is defined in the microcanonical ensemble, i.e.  $ \frac{1}{T}
\equiv \frac{\partial S}{\partial E}$. Eq.~\ref{eq:partial} can be derived using
the total differential $dF = dE - SdT -TdS$ combined with the
classical relation $S = -\frac{\partial F}{\partial T}$ in the
canonical ensemble.

 Using Eq.~\ref{eq:partial}, we can calculate the entropy $S(T)$ by
 integrating $1/T$ over the energy, i.e.
\begin{equation}
  \label{eq:S_int} S(T) = S(T_i) + \int_{E_i}^{E(T)} \frac{dE'}{T(E')}
\end{equation}
where $T(E)$ is the temperature associated to the average energy
$E(T)$ in Figure~\ref{fig:EvsT} and $T_i$ is the temperature at which
$E(T_i)=E_i$. To carry out this calculation we need to know the value
of the entropy $S(T_i)$ . For the liquid we choose the high
temperature limit as our reference state. At $T \rightarrow \infty$,
the entropy per spin is $S_\infty = \ln(2)$ (the factor $2$ reflecting
the $2$ possible states of any individual site) and the energy is the
average concentration of FLS's in a random state of the system, i.e.
\begin{equation}
  \label{eq:Einfty} E_\infty = -\frac{g}{2^6}
\end{equation}

\noindent where $g$ is the degeneracy of the FLS and is provided in
Figure~\ref{fig:crystal}. The quantity $S_{\infty}$ corresponds to the total size of the configuration space and so is independent of the choice of FLS. $E_{\infty}$, as given by Eq. ~\ref{eq:Einfty}, does depend on the FLS via the multiplicity $g$, That said, we note that $E_{\infty}$ is small in magnitude and, as we shall see, this dependence contributes little to the overall influence of the choice of FLS on the thermodynamics. For the crystal, the ground state energy $E_0$ (also
provided in Figure~\ref{fig:crystal}) provides the reference energy
with an entropy $S(T=0) = 0$.

\begin{figure}[p] \centering \vspace{-1.6cm}
  \includegraphics{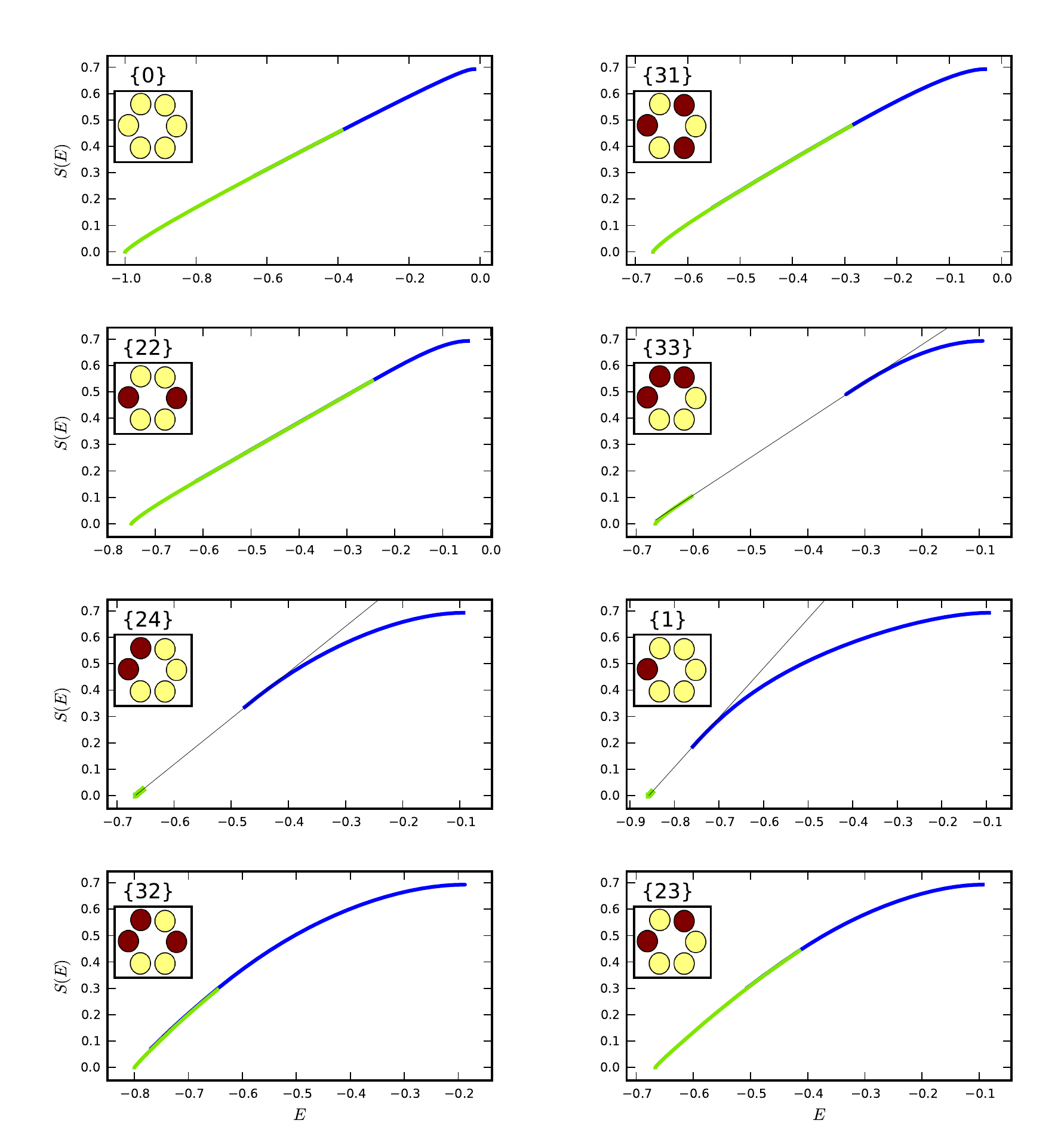}
  \caption{\label{fig:SvsE}(Color online) The entropy $S(E)$ for the
    crystal (green / light gray) and liquid (blue / dark gray) for
    each choice of the FLS. The common tangent is indicated as a thin
    black line in the case of first order transitions.}
\end{figure}

Using Eq.~\ref{eq:S_int} and the respective references states for the
liquid and solid, we have calculated the liquid and solid entropies by
numerical integration of the simulation data (in order to represent
correctly the high temperature liquid, we used a regular sampling in
the variable $1/T$ in this limit). In Figure~\ref{fig:SvsE} we plot
the entropy $S(T)$ against the energy $E(T)$ for the different choices
of FLS. As described above, the coexistence temperature $T_f$ is equal
to the inverse of the slope of the common tangent between the solid
and liquid entropy curves for those FLS's that exhibit a 1st order
freezing transition. These common tangents are indicated in each case
by a straight dashed line.  We have constructed this common tangent
with the help of fitted curves for the entropy (power law at low
temperature, polynomial for the liquid part), which serve only to get
rid of the noise for the numerical construction of the tangent. This
construction provides a precise estimate of the transition
temperature, even when hysteresis phenomena prevent us from reading it
directly on the $E(T)$ curve.

For continuous transitions, both Eq.~\ref{eq:helmholtz} and the
equivalent common tangent construction in the $S$-$E$ plane are
inadequate since there is no entropy nor energy difference between the
two phases at the transition. For these transitions, we determined the
transition temperature as the position of the peak in the heat
capacity. The values of $T_{f}$ for each of the choices of FLS
obtained either from the heat capacity peak or via the common tangent
construction are presented in Table~\ref{tab:Tf}.

\begin{table*}[h] \centering
  \begin{tabular}[c]{|c||c|c|c||c|c|c||c|c|}
\hline & \multicolumn{3}{|c||}{ Continuous } & \multicolumn{3}{|c||}{ First order} &\multicolumn{2}{|c|}{Multi-step} \\  \hline
 FLS & $\str{0}$ &$\str{31}$ & $\str{22}$ & $\str{33}$ & $\str{24}$  & $\str{1}$ & $\str{32}$ & $\str{23}$ \\
\hline
$T_{f}$ & 1.43 & 0.85 & 0.96 & 0.70  & 0.57 & 0.54 & 0.34 & 0.53 \\
 & & & & &  & & 0.50  & 0.57 \\
 & & & & & & & 0.56 & 0.64  \\ \hline
  \end{tabular}
  \caption{\label{tab:Tf} The freezing points of the various FLS
    systems. In the case of a continuous transition
    $T_{f}$ is determined as the
    position of the maximum of the heat capacity. In the case of a first
    order transition, $T_{f}$ is
    identified as the midpoint between the transition temperatures on heating and cooling. In case of multiple transitions, all the individual
    transition temperatures are presented. }
\end{table*}

The representation of the freezing transition in the $S$-$E$ plane
allows for the most transparent connection to be made between the
immediate consequences of the choice of FLS and the transition
temperature. To illustrate this connection, we shall consider the
following simple treatment. Let the liquid entropy be approximated by:
\begin{equation}
\label{gauss}
S(T)=S_{\infty}-A(E_{\infty}-E(T))^{2}
\end{equation}
As explained in the Appendix, Eq.~\ref{gauss} represents the first
term in an expansion of the liquid in terms of $(E_{\infty}-E)$ and
the constant $A$ can be evaluated exactly for this expansion. These
calculated values of $A$ are provided in ref.~\cite{ronceray_liq}. The
second assumption is to assume that the crystal entropy at coexistence
is zero. This assumption is justified by the observation on
Figure~\ref{fig:SvsE} that the entropy range for the crystal is indeed
very small as compared to the liquid's. A more accurate treatment of
the crystal entropy is provided by a single excitation model discussed
in the Appendix. With these two assumptions, the common tangent
construction is expressed by the relation

\begin{equation}
\label{approx}
2A(E_{\infty}-E)=\frac{S_{\infty}-A(E_{\infty}-E)^{2}}{E-E_{o}}
\end{equation}

\begin{figure}[ht] \centering
  \includegraphics{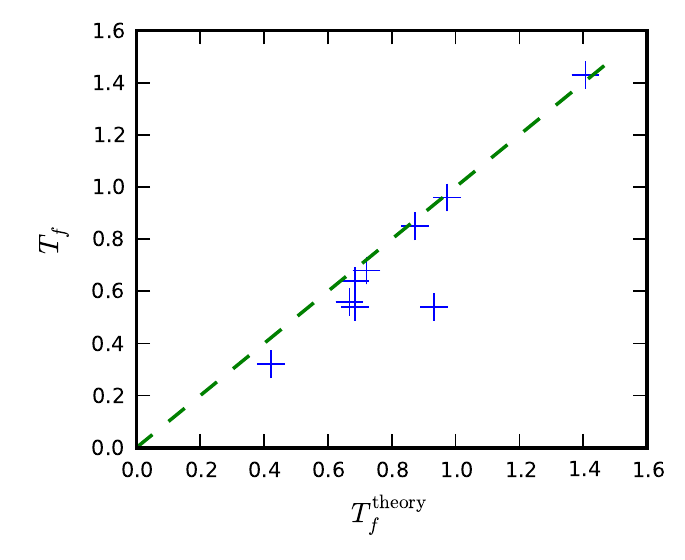}
  \caption{\label{fig:Ttheory} A scatter plot of the values of $T_f$ from simulation against the value of $T_{f}^{\mathrm{theory}}$ from Eq.~\ref{Tf}. The straight line with slope one indicates the case where $T_{f} = T_{f}^{\mathrm{theory}}$. }
\end{figure}

\noindent Solving for $E$, the value of the liquid energy at
coexistence, we have

\begin{equation}
\label{E}
E = E_{o}+\sqrt{\Delta^{2}-S_{\infty}/A}
\end{equation}
\noindent where $\Delta = E_{\infty}-E_{o}$. The theoretical estimate
of the transition temperature $T_{f}^{\mathrm{theory}}$, given by the inverse
slope of the common tangent, is

\begin{equation}
\label{Tf}
T_{f}^{\mathrm{theory}}=\frac{1}{2A(\Delta-\sqrt{\Delta^{2}-S_{\infty}/A})}
\end{equation}
\noindent As is evident from Eqs.~\ref{E} and \ref{Tf}, the freezing
transition is determined (within this approximate treatment) by just
two parameters: the crystal lattice energy, $E_{\infty}-E_{o}$, and
the liquid entropy factor $A$. (The high temperature properties $S_{\infty}$ and $E_{\infty}$ being independent of FLS or approximately so, respectively.) The factor $A$ contains the entropy
cost of structure in the liquid : a large value of $A$, generally
associated with a high symmetry FLS, corresponds to a local structure
that incurs a large entropy cost. In Fig.~\ref{fig:Ttheory} we plot
the values of $T_{f}^{\mathrm{theory}}$ calculated using Eq.~\ref{Tf} with the
values of the freezing temperature from Table~\ref{tab:Tf}. Given the
simplicity of the treatment, the theoretical estimates of the freezing
temperatures are quite reasonable. The notable exception to this
success is the FLS \str{1} where the theory considerably overestimates
transition temperature. The \str{1} liquid actually exhibits a
substantial decrease in energy prior to freezing, which is not
correctly taken into account by the factor $A$, as previously
noticed~\cite{ronceray_liq}, which explains this deviation. While this
approximate treatment is too simple to establish whether a transition is
first- or second-order, we note that it does appear to get the
magnitude of the transition temperature roughly right, even for
continuous transitions.

\section{Freezing in Multiple Steps}

\begin{figure}[ht] \centering
  \includegraphics{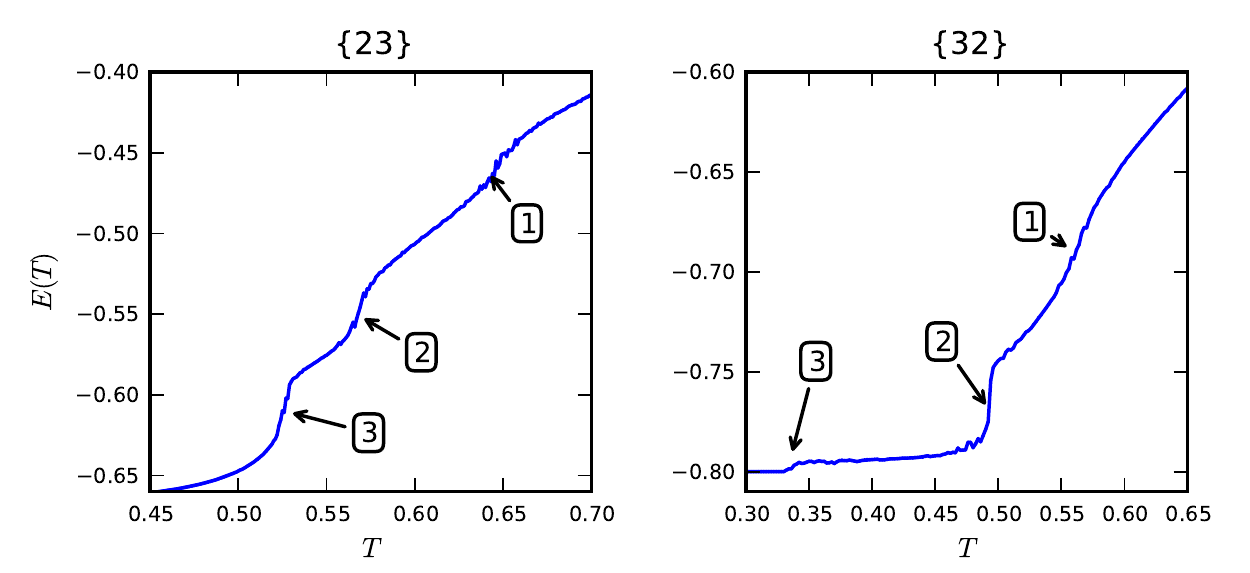}
  \caption{\label{fig:mult_trans} Energy versus temperature
curve for multiple transitions systems. The positions of
the individual transitions are indicated. }
\end{figure}

The presence of multiple transitions for the \str{23} and \str{32}
FLS's is clearly seen when we examine the $E(T)$ curves with an
expanded temperature scale about the transition region, as shown
in Figure~\ref{fig:mult_trans}. In both cases, the freezing transition
is made up of three distinct transitions. This is unusual for
crystallization which is typically viewed as the archetype of a highly
cooperative transition and hence unlikely to decouple in this way. In
this Section we shall try to identify the intermediate phases
associated with the step-wise freezing processes.

In the case of the \str{23} FLS, the multiple transition scenario
appears to arise from the existence of the two degenerate polymorphs
(see Figure~\ref{fig:crystal}). If we identify the three transition
temperature as $T_1$, $T_2$ and $T_3$, as indicated on
Figure~\ref{fig:mult_trans}, it appears that the transition on cooling
through $T_2$ corresponds to the formation of a polycrystalline
mixture of the two polymorphs. Further cooling sees the formation at
$T_3$ of the single phase of the preferred crystal phase. The nature
of the continuous transition at $T_1$ remains a puzzle.

Turning to the transitions in the \str{32} system , the picture is
quite different. Instead of different crystal forms, the sequence of
transitions correspond to a step-by-step symmetry breaking.
Transition 1 marks the appearance of an orientation, at which the
symmetry of the liquid is reduced from ``isotropic'' to a liquid with
3-fold rotational symmetry. On cooling, transition 2 is characterised
by the loss of this rotational symmetry as a unique orientation is
selected.  Finally, transition 3 breaks the reflection symmetry and
the translational invariance in one direction, to give the low $T$
crystalline state. Transitions 2 and 3 seem to be first order, though
the small energy gap at transition 3 permits oscillation between
phases in our finite size systems. This sequence of symmetry breakings
is reminiscent of the liquid-nematic-smectic sequence of transitions
observed in some molecular liquids.

\section{Discussion}

In the case of a single FLS, the choice of the stable structure
determines the freezing transition temperature by determining the
energy of the crystal lattice and the entropy cost of the FLS in the
liquid. How might this picture change when we shift our focus to model
liquids in which variations of the particle interactions are used to
alter the local favoured structure? Previous work on modifications of
the inter-particle potential for tetrahedral particles provides some
insight into how subtle this problem can be.  In the study by Molinero
{\it et al}~\cite{sastry} on the modified silicon potential,
increasing the strength of the 3-body bending force constant results
in an increase in the temperature at which the liquid freezes into the
diamond structure. Since the bending force constant has little effect
on the solid, the influence of the potential acts through the
liquid. An analysis of the enthalpy data~\cite{sastry-supp} for the
modified silicon liquid indicates that the major effect of the
increase in the bending constant is to increase the liquid enthalpy,
an effect absent from the FLS model. In a model of spheres with a
tetrahedral arrangement of sticky patches, the angular width $\sigma$
(in radians) of the individual patch represents another parameter for
controlling the flexibility of the local structure. Doye et
al~\cite{doye}have found that this model fails to crystallize at all,
arguing that, for patch sizes down to 0.2 radians, there are multiple
near degenerate structures - dodecahedral clusters, cubic diamond and
hexagonal diamond - with the result being some sort of structural
'traffic jam'. In neither example cited here has the modification of
the inter-particle potential corresponded to a change in the local
geometry.

\section{Conclusion}

In this paper we have examined the variety of freezing transitions
exhibited by a model liquid characterised by a favoured local
structure. We found examples of ordering transitions that were first
order, continuous and multi-stepped. All liquids were found to order
with little or, in the case of the continuous transitions, no evidence
of a tendency to supercool, in spite of the involvement of some
complex crystal phases. The variation of the transition temperature with choice of favoured local structure was found to be reasonably captured by a simple expression involving the groundstate energy $E_{o}$ and the factor $A$ related to the entropy cost per FLS in the liquid. 

A couple of observations follow directly from the results reported
here. i) As all choices of FLS, with the exception of \str{0}, are
frustrated (i.e. not all sites can lie simultaneously in the FLS :
$E_{o} > -1$), it is clear that frustration provides no inherent
impediment to freezing and is generally insufficient to stabilize the
supercooled liquid. The competition between {\it two} FLS's: one that
can produce a low energy crystal while the other FLS, although more
stable, can only order in a higher energy structure, looks like the
simplest means of generating a glass forming liquid. ii) The intuitive
notion that large unit cell crystals are more 'difficult' to
crystallize finds little support here. We note that the very
feature (\emph{i.e }low symmetry of the FLS) that results in large unit cells
also ensures that the liquid, prior to freezing, can accumulate a
substantial amount of the favoured local structure. As we have seen,
these two effects tend to counter one another in setting the value of
the transition temperature.

The FLS model provides a useful tool for building up our intuition on how
local structure influences the properties of condensed matter. A
report on the extension to 3D is currently in
preparation~\cite{ronceray-fcc}. By building the model around the idea of
local stable structures, rather than having to discover them as a
consequence of particle interactions, we believe that the FLS approach
offers a clearer language with which to ask questions about structure
in condensed phases - both the influence of local structure and the
nature of global ordering.

{\bf Acknowledgements}

We gratefully acknowledge funding from the \'{E}cole Normale
Sup\'{e}rieure and the Australian Research Council.

\vspace{30pt}

{\bf APPENDIX}

In this Appendix we describe an alternative to the numerical fits used
for the common tangent construction, which gives analytic values of
the Taylor expansion coefficients around asymptotic high- and
low-temperature regimes. We formerly derive the first order term of
the expansion~\cite{ronceray_liq}. Here we generalize these results to
any order using correlation functions that can be computed by exact
enumeration. We also present a route to approximate the thermodynamic
quantities at low temperature, though only to leading order.

\subsection{The Energy of the Low Temperature Crystal }

\begin{figure}[p] \centering 
  \includegraphics[width = 0.7\textwidth]{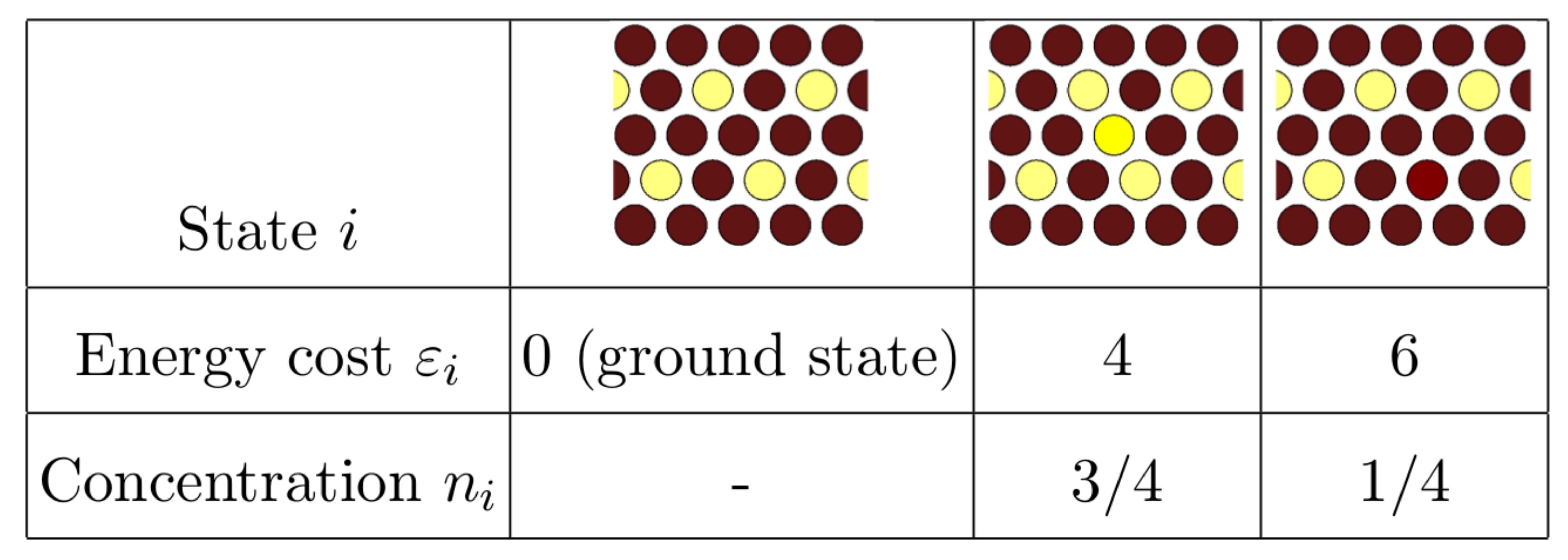}
  \caption{\label{tab:excitations} (Color online) Sketch of the energy spectrum of
    system \str{22} at low temperature. For this system, two kinds of
    elementary (single spin) defects can occur in the crystal, examples of
    which are presented schematically (the ground state is also
    reminded). The energy spectrum of a crystal is determined by the
    energy costs $\varepsilon_i$ of each kind of defect, and the
    respective concentration of sites that can host such a defect $n_i$.}
\end{figure}

We shall treat the energy of the crystal phase using a low temperature
approximation in which we only include the (uncorrelated) single spin
excitations from the groundstate. First, we identify the energy
spectrum $(n_i,\varepsilon_i)$ of the ground state, as introduced in
Table~\ref{tab:excitations} in the example of system \str{22}. It
classifies the single-spin defects that can occur in the crystal
according to the energy of the corresponding state $\varepsilon_i$ and
the proportion of site that can host such a defect $n_i$.

In the canonical ensemble at temperature $T$, each site of type $i$
has a probability:
$$p_i(T) = \frac{e^{-\varepsilon_i/T}}{1+e^{-\varepsilon_i/T}}$$
to be in its excited state, if we negligate the interaction between
such defects. This gives the following approximate expression for the
energy :
\begin{equation}
  \label{eq:low_T_model} E(T) = E_0 + \sum_i \varepsilon_i n_i
\frac{e^{-\varepsilon_i/T}}{1+e^{-\varepsilon_i/T}}
\end{equation}
All other thermodynamic quantities derive from $E(T)$. However the
computation of $S(T)$, for example, involves complicated analytic
forms, and it is much easier to compute it by numerical integration of
Eq.\ref{eq:low_T_model} using Eq. \ref{eq:S_int}, to arbitrary
precision.

\begin{figure}[ht] \centering
   \includegraphics{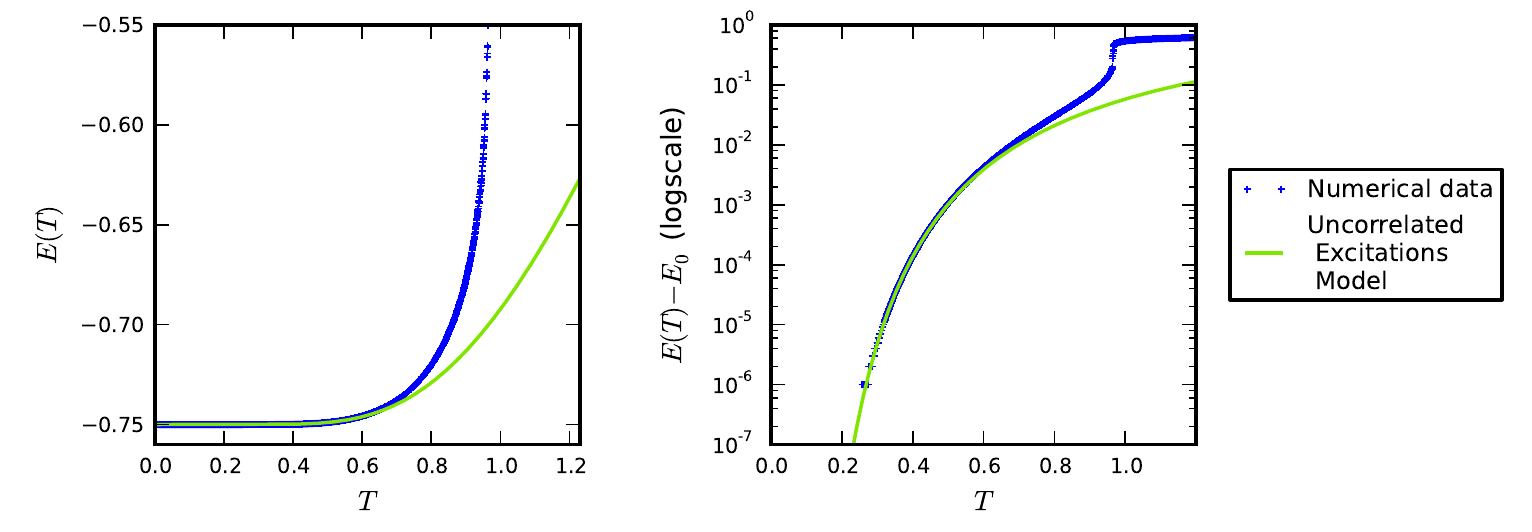}
   \caption{\label{fig:low_T_model}(Color online) The low temperature
     model described in equation~\ref{eq:low_T_model}, compared to
     simulated results for system \str{22}. The agreement is excellent
     at low temperature, as can be seen on the right figure, which is
     $E(T)-E_0$ plotted in logarithmic scale. This model correctly
     identifies leading-order behaviour at low temperature.}
\end{figure}

A quantitative comparison between the results of simulations and this
model is presented on Figure~\ref{fig:low_T_model}, for system
\str{22}. We find that the accordance is quantitative up to a
temperature around $T=0.6$. For the other systems, no qualitative
difference was observed, though the quantitative values in the energy
spectrum vary. This simple model gives a satisfying description of the
crystal state, the physics of which is therefore quite simple. This
method gives easily the leading order of the low-temperature
expansion. Finding the corrections to this expansion would involve
correlation between defects. No difficulty is expected to arise from
that, though the computations would be rather tedious.

\subsection{Exact high-temperature expansion of the FLS model at all orders}

In this section, we present an adapted version of the classical
high-temperature expansions of spin models adapted to the FLS
model. It gives the exact values of the coefficients of the Taylor
expansion of the thermodynamic functions around $\beta = 1/T =
0$. These coefficients are connected correlation functions at infinite
temperature, which may be computed exactly. This reasoning is an
extension of our work in~\cite{ronceray_liq}, where we presented
the route to compute the first non-trivial coefficient.

\subsubsection{Definitions :}

To make the link with other spin models, we first introduce the
Hamiltonian (energy of a configuration) :
\begin{equation}
  \label{eq:H}
  H(\mathcal{C}) = \sum_{i\in\Lambda} \epsilon_i
\end{equation}
where $\Lambda$ is the lattice, $\mathcal{C}$ the spin configuration,
and the local energy variables are :
\begin{equation}
  \label{eq:ei}
  \epsilon_i =
  \begin{cases}
    -1 & \text{ if $i$ is in the FLS in configuration }\mathcal{C} \\
    0 & \text{otherwise}
  \end{cases}
\end{equation}

Each site of the lattice has $z$ neighbours ($z$ is the coordinence of
the lattice). We assume invariance under translations, which impose to
use periodic boundary conditions if the system is of finite size. For
the triangular lattice studied here $z=6$. Note that the FLS model
could be adapted to other regular lattices ; a review of this model on
the 3D FCC lattice is in preparation.

The partition function of the system is defined as :
\begin{equation}
  \label{eq:Z}
  \mathcal{Z}(\beta) = \sum_\mathcal{C} \exp(-\beta H(\mathcal{C}))
\end{equation}
where $\beta = 1/T$ is the inverse temperature. All thermodynamic
quantities derive from $\mathcal{Z}$ and its derivatives ; for example
:
\begin{eqnarray}
  F(\beta) = &-\frac{1}{\beta} \log\mathcal{Z}(\beta) \\
  E(\beta) = &-\frac{1}{\mathcal{Z}(\beta)} \frac{\partial \mathcal{Z}(\beta)}{\partial \beta} \\
  S(\beta) = & \beta( E - F)
\end{eqnarray}
are respectively the free energy, the average energy and the entropy
of the system.

We define an average $\av{\cdot}_0$ as the uniform average over all
configurations, \emph{i.e.} at $\beta = 0$.

Note that for any site $i$, we have :
\begin{equation}
  \label{eq:p}
  \av{\epsilon_i}_0 = E_\infty = -\frac{g}{2^z}
\end{equation}
where $g$ is the \emph{degeneracy} of the FLS and $E_\infty$ is the
average energy per site at infinite temperature.

\subsubsection{Derivation}

Now that the notations are set, we are ready to derive an exact Taylor
expansion, at any order, of the thermodynamic functions in the FLS
model around $\beta = 0$.

Working on the partition function $\mathcal{Z}(\beta)$, a few usual
tricks allow us to rewrite:
\begin{eqnarray}
  \label{eq:play_with_Z}
  \mathcal{Z}(\beta) &= \sum_\mathcal{C} \exp(-\beta H(\mathcal{C})) \\
  &= \sum_\mathcal{C} \prod_{i\in\Lambda} \exp(\beta \epsilon_i ) \\
  &= \sum_\mathcal{C} \prod_{i\in\Lambda} ( 1 + \nu  \epsilon_i )
\end{eqnarray}
with $\nu = e^\beta -1$. Note that $\nu \approx \beta$ at high
temperature. These simplifications use the fact that $\epsilon_i$
takes only two values, $0$ and $1$.

Now we can expand this product exactly :
\begin{equation}
  \prod_{i\in\Lambda} ( 1 + \nu  \epsilon_i ) = \sum_{A\subseteq \Lambda}  \nu^{|A|} \prod_{i\in A}  \epsilon_i
\end{equation}
The meaning of this expansion is the following : $A$ is the subset of
right-side factors selected in the expansion of the product, and $|A|$
is the cardinal of $A$. If we now sum over configurations
$\mathcal{C}$ :
\begin{eqnarray}
    \mathcal{Z}(\beta) = &  \sum_\mathcal{C} \prod_{i\in\Lambda} ( 1 + \nu  \epsilon_i ) \\
    = & 2^N \sum_{A\subseteq \Lambda} \nu^{|A|}\av{\prod_{i\in A}  \epsilon_i}_0
\end{eqnarray}
where $N$ is the total number of sites. We can rearrange this sum
according to $|A|$ :
\begin{equation}
  \mathcal{Z}(\beta) = 2^N \sum_{k\in\mathbf{N}} \sum_{A\subseteq \Lambda,\ |A| = k} \nu^k\av{\prod_{i\in A}  \epsilon_i}_0
\end{equation}

So finally we get the high temperature explicit, exact expansion :
\begin{equation}\boxed{
  \mathcal{Z}(\beta) = 2^N \sum_{k\in\mathbf{N}}\ \ \sum_{i_1 \neq \dots \neq i_k} \nu^k\av{ \epsilon_{i_1}  \epsilon_{i_2} \dots  \epsilon_{i_k}}_0}
\end{equation}

A convenient rewriting of this equation in terms of the free energy
leads us to introduce the connected correlation functions:
\begin{equation}\boxed{
  \beta F(\beta) = -\log \mathcal{Z}(\beta) = -N\log(2) - \sum_{k>0}\ \ \sum_{i_1 \neq \dots \neq i_k} \nu^k\av{ \epsilon_{i_1}  \epsilon_{i_2} \dots  \epsilon_{i_k}}_0^C}
\end{equation}
where :
\begin{align*}
  \av{\epsilon_i}_0^C &= \av{\epsilon_i}_0 \\
  \av{\epsilon_i\epsilon_j}_0^C &= \av{\epsilon_i \epsilon_j}_0 - \av{\epsilon_i}_0\av{\epsilon_j}_0 \\
  \av{\epsilon_i\epsilon_j\epsilon_k}_0^C &= \av{\epsilon_i \epsilon_j\epsilon_k}_0 - \av{\epsilon_i}_0\av{\epsilon_j\epsilon_k}_0 - \av{\epsilon_j}_0\av{\epsilon_i\epsilon_k}_0 - \av{\epsilon_k}_0\av{\epsilon_i\epsilon_j}_0 + 2 \av{\epsilon_i}_0\av{\epsilon_j}_0\av{\epsilon_k}_0 \\
  \dots
\end{align*}

For the explicit computation, free energy representation is more
convenient since connected correlation functions vanish as soon as a
site is ``far'' (not overlapping) from the others. Thus every term in
the expansion is extensive in the size of the system. In order to
compute explicitly the connected correlation function at order $k$, we
need to enumerate all possible ways to put an FLS in each of the $k$
sites $i_1 \dots i_k$. It will be zero if they can be separated in two
non-overlapping set of structures, thus we need only to enumerate a
finite number of positions $i_1\dots i_k$. The quantity $\av{
  \epsilon_{i_1} \epsilon_{i_2} \dots \epsilon_{i_k}}_0$ can be
interpreted simply as the probability that all sites $i_1 \dots i_k$
lie simultaneously in the FLS, in a random spin configuration.

\subsubsection{Computation of the first terms}

The first terms read :
\begin{equation}
\label{eq:free-energy}
  \beta F(\beta) = -N\log(2) - \nu \sum_{i\in\Lambda} \av{\epsilon_i}_0^C + \nu^2 \sum_{i\neq j} \av{\epsilon_i \epsilon_j}_0^C + \nu^3 \sum_{i\neq j\neq k} \av{\epsilon_i \epsilon_j \epsilon_k}_0^C + O(\nu^4)
\end{equation}
The truncature to the order $\nu^2$ would give the Gaussian
approximation to the density of states, as in~\cite{ronceray_liq}. Now
the interest of this new, more formal approach, is to be able to
increase the precision and involve higher-order correlation functions.

All that is left is to compute explicitly these connected correlation
functions, which need only to test a finite (but possibly big) number
of configurations. This is the only model-specific part of this
derivation. Using enumerating routines, we could quite easily
get the $\nu^3$ term for FCC and triangular lattices.  Finally, to
obtain an expansion of the free energy in powers of $\beta$ instead of $\nu$, we can substitute the expansion of $\nu$ in terms of $\beta$, i.e.
\begin{equation}
  \nu = e^\beta - 1 = \sum_{k>0} \frac{\beta^k}{k!} = \beta + \beta^2/2 + \beta^3/6 + \dots
\end{equation}

\noindent into Eq.~\ref{eq:free-energy}.

\end{document}